\documentclass{article}
\usepackage{natbib,epsfig,emulateapj}

\def\be{\begin{equation}}
\def\ee{\end{equation}}
\def\ba{\begin{eqnarray}}
\def\ea{\end{eqnarray}}
\def\go{\mathrel{\raise.3ex\hbox{$>$}\mkern-14mu
             \lower0.6ex\hbox{$\sim$}}}
\def\lo{\mathrel{\raise.3ex\hbox{$<$}\mkern-14mu
             \lower0.6ex\hbox{$\sim$}}}
\def\etal{{\it et al.~}}

\begin{document}

\title{Pulsar Jets: Implications for Neutron Star Kicks and
Initial Spins}
\author{Dong Lai, David F. Chernoff, and James M. Cordes}
\affil{Center for Radiophysics and Space Research, Department of
Astronomy, Cornell University,
Ithaca, NY 14853\\
dong,chernoff,cordes@spacenet.tn.cornell.edu}

\begin{abstract}
We study implications for the apparent alignment of the spin axes,
proper-motion directions, and polarization vectors of the Crab and
Vela pulsars. The spin axes are deduced from recent Chandra X-ray
Observatory images that reveal jets and nebular structure having
definite symmetry axes. The alignments indicate these pulsars were
born either in isolation or with negligible velocity contributions
from binary motions. We examine the effects of rotation and the
conditions under which spin-kick alignment is produced for theoretical
models of neutron star kicks.  If the kick is generated promptly
during the formation of the neutron star by asymmetric mass ejection
and/or neutrino emission, then the alignment requires that the
protoneutron star possess, by virtue of the precollapse stellar core's
spin, an original spin with period $P_s$ much less than the kick
timescale $\tau_{\rm kick}$, thus spin-averaging the kick forces on
the star.  The kick timescale ranges from $100$~ms to $10$ seconds
depending on whether the kick is hydrodynamically driven or
neutrino-magnetic field driven. For hydrodynamical models, spin-kick
alignment further requires the rotation period of an asymmetry pattern
at the radius near shock breakout ($\go 100$~km) to be much less than
$\tau_{\rm kick}\lo 100$~ms; this is difficult to satisfy unless
rotation plays a dynamically important role in the core collapse and
explosion (corresponding to $P_s\lo 1$~ms). Aligned kick and spin
vectors are inherent to the slow process of asymmetric electromagnetic
radiation from an off-centered magnetic dipole. We reassess the
viability of this electromagnetic rocket effect, correcting a factor
of 4 error in Harrison and Tademaru's calculation that increases the
size of the effect. To produce a kick velocity of order a few hundred
km~s$^{-1}$ requires that the neutron star be born with $P_s \sim 1$ ms
and that spindown due to r-mode driven
gravitational radiation be inefficient compared to standard magnetic
braking.  
The electromagnetic rocket operates on a timescale
of order $0.3\,(B/10^{13}\,{\rm G})^{-2}$~year.
The apparent spin-kick alignment in the Crab and Vela
pulsars places important new constraints on each of the mechanisms of
neutron star kicks we consider.

\end{abstract}


%%%%%%%%%%%%%%%%%%%%%%%%%%%%%%%%%%%%%%%%%%%%%%%%%%%%%%%%%%%%%%%%%%%
\section{Introduction}

It has long been recognized that neutron stars 
have space velocities much greater than those of their progenitors.
% (e.g., Gunn \& Ostriker 1970). 
Recent studies of pulsar proper motion give $200-500$~km~s$^{-1}$ as the mean
3D velocity of neutron stars at birth (e.g., Lyne and Lorimer 1994; 
Lorimer et al.~1997; Hansen \& Phinney 1997; Cordes \& Chernoff 1998), and a
significant population having velocities greater than
$1000$~km~s$^{-1}$. Direct evidence
for pulsar velocities $\go 1000$~km~s$^{-1}$ comes from observations of 
the bow shock produced by the Guitar Nebula pulsar (B2224+65) 
in the interstellar medium (Cordes, Romani \& Lundgren 1993). 
A natural explanation for such high velocities is that supernova 
explosions are asymmetric, and provide kicks to nascent neutron stars. 
%Support for kicks has come from the detection 
The geodetic precession in the binary pulsar PSR 1913+16 
(Weisberg, Romani \& Taylor 1989; Cordes et al.~1990; Kramer 1998; 
Wex et al.~2000), the orbital plane
precession in PSR J0045-7319/B-star binary and its fast 
orbital decay (Kaspi et al.~1996; Lai et al.~1995; Lai 1996),
and the high eccentricities of Be/X-ray binaries (see van den Heuvel \& van
Paradijs 1997) all support the notion of neutron star kicks. 
The recently detected high systemic velocity ($\simeq
430$~km~s$^{-1}$) of the X-ray binary Circinus X-1 (Johnston et al.~1999) can
only be produced by a neutron star kick of at least $\sim 500$~km~s$^{-1}$
(Tauris et al.~1999). Evolutionary studies of neutron star binary population 
also imply the existence of pulsar kicks (e.g., Dewey \& Cordes 1987; 
Fryer \& Kalogera 1997; 
Fryer et al.~1998). Finally, there are many direct
observations of nearby supernovae (e.g., Wang et al.~1999) and supernova
remnants which show that supernova explosions are not spherically
symmetric.

While the evidence for such kicks is unequivocal,
the physical origin remains unclear.
A natal kick could be generated by an asymmetric explosion due to global
hydrodynamical perturbations in the supernova core 
(Goldreich, Lai \& Sahrling 1996; Burrows \& Hayes 1996; Lai 1999;
Lai \& Goldreich 2000a,b), or it could be a result of asymmetric neutrino
emission in the presence of superstrong magnetic fields ($B\go 10^{15}$~G) 
in the proto-neutron star (e.g., Lai \& Qian 1998; Arras \& Lai 1999a,b and
references therein). A post-natal kick mechanism studied 
by Harrison and Tademaru (1975) relies on asymmetric electromagnetic
radiation from an off-centered dipole in a rapidly rotating pulsar.
All these possibilities have intrinsic uncertainties 
(see \S 3 and \S 4 below).  
Part of the reason that it is difficult to determine which mechanism
is at work is the lack of any simple relationship between
velocity and other properties of neutron stars (but see \S 2). For example,
despite some early claims, statistically there is no evidence
for a correlation between velocity and magnetic moment for radio pulsars
(Lorimer et al. 1997; Cordes \& Chernoff 1998),
or between velocity and rotation 
(Deshpande, Ramachandran \& Radhakrishnan 1999).
Unfortunately, given the large systematic uncertainties, these 
statistical results, by themselves, can not be used to
support or rule out particular kick mechanisms (see \S 2.3).
 
Recent observations of the Vela pulsar and the surrounding compact X-ray nebula
with the Chandra X-ray Observatory reveal a two sided asymmetric jet
at a position angle coinciding with the position angle of the pulsar's
proper motion (Pavlov et al.~2000). The symmetric morphology of the nebula
with respect to the jet direction strongly suggests that the jet is
along the pulsar's spin axis. In \S2.1 we examine the polarization angle
of Vela's radio emission which corroborates this 
interpretation. Evidence for spin-velocity alignment also exists
for the Crab pulsar (\S 2.2). We argue in \S 2.3 that current 
observations of other pulsars neither support nor rule out any 
spin-kick correlation.

The alignment between the projected spin axis and proper motion (for Vela
and Crab) immediately demonstrates that binary break-up along with 
symmetric supernova explosions 
%(as originally suggested by Gott et al.~1970) 
(see Iben \& Tutukov 1996) is not
likely to be responsible for the observed 
proper motion. Binary disruption would yield orthogonal spin and 
velocity vectors if pre-explosion binaries have aligned spin and orbital
angular momenta (e.g., by virtue of tidal coupling) and if the
neutron star's spin is in the same direction as the progenitor's.  Projection
onto the plane of the sky could fortuitously yield apparently parallel
spin and velocity vectors, of course, but this is a low probability 
occurrence.   
The observed spin-velocity alignment in Vela and Crab pulsars then indicates 
that the pulsar was born in isolation, or the progenitor binary was 
so wide that the orbital speed was negligible compared to the kick.
In \S 3 and \S4 below, we ignore the binary possibility, and 
examine how rotation may affect kick and under what conditions
spin-kick alignment is expected for different kick mechanisms. 
%explore the implications of the observed alignment for
%the kick mechanisms and related issues.  

%%%%%%%%%%%%%%%%%%%%%%%%%%%%%%%%%%%%%%%%%%%%%%%%%%%%%%%%%%%%%%%%%%%
\section{Observational Constraints on the Geometry of Pulsar Spin and Velocity}

Here we briefly discuss observational results related to 
spin -- proper motion correlation in pulsars.

\subsection{Vela Pulsar}

The proper motion of the Vela pulsar 
$(\mu_{\alpha}, \mu_{\delta}) = (-48\pm 2, 35\pm 1)$ mas~yr$^{-1}$ 
(Bailes et al.~1990) is at position angle $\psi_{\mu} 
= -53.9^\circ\pm 1.4^\circ$ and implies a 2D velocity $D\mu \simeq 70-141$
km~s$^{-1}$ for a distance
$D=0.25-0.5$ kpc. The proximity of the pulsar to the Earth
implies that any correction
for differential galactic rotation is small.
The X-ray jet and bow-shock axes of the pulsar nebula lie in the
same direction as the proper motion to within 5 degrees (Pavlov et al.~2000).
If the jet originates from the pulsar magnetosphere, as seems likely,
it is most natural to associate the jet axis with the pulsar spin axis.

Another constraint on the geometry comes from radio polarization.
The Vela pulsar formed the basis
for the rotating vector model (Radhakrishnan \& Cooke 1969), which 
relates the polarization position angle to the rotating dipolar magnetic
field of the neutron star. At the pulse centroid, the spin axis,
magnetic moment and line of sight are coplanar. 
Because the radiation is highly beamed, 
the polarization angle at the pulse center is determined by the 
orientations of the stellar magnetic field and
of the projected spin axis.  Phenomenologically, radio emission
often conforms to this picture, but the polarization angle
can be in one of two orientations separated by 90 degrees. 
Among various pulsars, the two modes
vary stochastically from pulse to pulse as well as systematically
across pulse phase.  In some objects, the emission seems to be dominated
by one mode in all pulses and at all pulse phases.    

Using data from Deshpande et al.~(1999),
which have been corrected for Faraday rotation in the interstellar medium
and in the ionosphere, the pulse centroid polarization angle
is $\psi_{\rm pol} = 35^\circ\pm 10^\circ$.  The difference angle between
the proper motion and polarization angle is 
$\Delta\psi = \psi_{\rm pol} - \psi_{\mu} = 89^\circ\pm 11^\circ$,
i.e., the polarization vector is perpendicular to the proper motion axis.  

We conclude that the space velocity and X-ray jet
axes are parallel in projection on the plane of the sky.  Moreover,
the radio polarization angle is 90 degrees with respect to this axis,
implying that the polarization mode dominant in the Vela pulsar is one
where the electric field in radio emission is orthogonal to the
magnetospheric field.

The {\it a priori} chance of alignment (or anti-alignment) of three
direction vectors (corresponding to proper motion, perpendicular
polarization, and X-ray jet axes) is small.  We assume each is
uniformly distributed on a sphere and use measurement uncertainties of
1.4, 10, and 5 degrees, respectively to infer a probability of 0.6\%.
The corresponding probability increases to 1.2\% if alignment of
either the parallel or the perpendicular polarizations are considered.

\subsection{Crab Pulsar}

Caraveo \& Mignani (1999) report a new HST-derived proper motion for
the Crab pulsar, $(\mu_{\alpha},\mu_\delta) = (-17\pm 3, 7\pm3)$ mas~yr$^{-1}$,
or $\mu = 18\pm3$ mas~yr$^{-1}$ with a position angle $\psi_\mu 
= 292\pm10^{\circ}$, corresponding to a transverse velocity of 171~km~s$^{-1}$
for $D=2$~kpc. Their results are consistent with those reported by Wyckoff \&
Murray (1977) using historical images of the Crab Nebula.
Though $\sim 2$ kpc away, differential galactic rotation is negligible
because the pulsar is within $5^{\circ}$ of the Galactic
anticenter direction.
Caraveo \& Mignani pointed out that the proper motion and the symmetry axis 
of the inner X-ray nebula, based on ROSAT images, were essentially parallel.  
Arguing that the X-ray structure's symmetry is determined by the pulsar's
spin axis, they suggested that the (projected) spin axis and proper motion
are parallel. The recent Chandra X-ray image (see Weisskopf 1999)
confirms this picture which is also consistent with
the optical work of Hester et al.~(1995). While the morphology of the 
X-ray emission is less structured and less symmetric than the Vela pulsar's
nebula, there is a well-defined axis that indeed is parallel with the 
proper motion, to within the errors.   

Interpretation of the Crab pulsar's polarization is less clear than that of 
Vela, primarily because there is a multiplicity of components 
whose relationship to the spin axis is ambiguous. However, radio polarization
at 1.4 GHz (Moffett \& Hankins 1999) has position angle (after correction 
for Faraday rotation) $\psi_{\rm pol} \simeq -60^{\circ}\pm10^{\circ}$ for
both the main pulse (MP) and interpulse (IP) components. At 5 GHz, the
polarization angle for the MP is the same while the IP angle is shifted by
$90^{\circ}$ to $30^{\circ}\pm10^{\circ}$.  At 8 GHz, the MP is not visible
while the IP has the same position angle as at 5 GHz.
Two new radio components (HFC1, HFC2) appear at 8 GHz 
that arrive later than the IP and have position angles 
$\sim 85^{\circ}\pm15^{\circ}$. At frequencies less than 1 GHz, a `precursor'
component appears before the MP but its polarization angle is not available.
At optical and UV wavelengths (Graham-Smith et al 1998, 1996), 
the polarization angles at the centroids
of the MP and IP are the same as in the radio (modulo $90^{\circ}$
polarization-mode ambiguities).

Empirically, the polarization vector at the centroid of the MP is
parallel to the proper motion vector. The simplest interpretation
is that the polarization mode in this pulse component is parallel to
the projected spin axis.
% as with the Vela pulsar.  
However, alternative models exist for the Crab pulsar's emission that place the
emission region at sufficiently high altitude (in `outer-gap' models) that
rotational aberration would break the coplanarity of the line of sight, spin
and magnetic dipole vectors (e.g. Romani \& Yadigaroglu 1995).
The geometry of the Crab's X-ray nebula may help clarify the emission 
geometry.

\subsection{The NS-NS Binary, B1913+16}

Analysis of the orbital elements, proper motion, and limits on the
geodetic-precession angle leads to the conclusion that a kick resulted
from the most-recent supernova that produced the observed pulsar's
companion and constrains the angle between the kick and orbital plane
(Wex \etal 2000). The angle is $<5-10$ degrees when the following
hold: (1) all angular momenta (spin and orbital) were aligned prior to
the explosion (consistent with the canonical evolutionary history of
double neutron star systems), (2) the kick timescale was much shorter
than the orbital period (currently $\sim 8$ hr), and (3) the orbit did
not evolve significantly since the explosion (see Fryer \& Kalogera
1997). If the current determination of the hard-to-measure geodetic
angle is ignored then the angle between kick and orbital plane is
limited to $<30$ degrees primarily by the condition that the binary
remain bound.

For all such analyses the following is important: the axis of the spin
of the most-recently formed neutron star is unobserved, so the
spin-kick orientation can be inferred only if one also assumes the
neutron star spin was aligned with the progenitor's spin. Although
plausible, there is no proof that the spin direction of the iron core
of this neutron star progenitor matched the presupernova orbital
angular momentum. We conclude that the results for B1913+16 do not
demand a different sort of kick mechanism than for the Crab and Vela
pulsars. At the same time, if spin-kick alignment holds for B1913+16,
if the angle between kick and orbital plane is $<30$ degrees and if
all angular momenta align, then the spin of the most-recently formed
neutron star in that system is misaligned with the spin of the
progenitor.

\subsection{Other Pulsars}
\label{sec:others}

The alignments of proper motion vectors and projected spin axes for
the Vela and Crab pulsars may be contrasted with what we know about
other pulsars' geometries. For other pulsars, we have no information
about the orientation of the spin axis from nebular morphology.  All
that is known is based on radio polarization.  For the 28 pulsars
(besides Vela) analyzed by Deshpande et al.~(1999), there appears to
be no systematic relationship between the polarization and proper
motion vectors. We have used a likelihood analysis to compare
different model relationships between these vectors. The null
hypothesis, that they are independently and uniformly distributed on
the sphere, was compared with two competing hypotheses: (1) that there
is perfect correlation between the spin axis and either one
polarization mode or its orthogonal counterpart; (2) a hybrid model
where some objects have uncorrelated angles and others have perfectly
correlated angles, modulo the 90 degree ambiguity.  The likelihood
analysis indicates, formally, that the best model is one where 80\% of
the objects have uncorrelated angles while 20\% have deterministically
related angles.  However, the likelihood for this model is not
significantly better than the null hypothesis. For the population of
objects, we conclude, therefore, that there is no strong relationship
between the polarization angle and the orientation of the proper
motion.

The lack of any empirically establishable correlation for
a sample of pulsars should not be viewed
as evidence that kicks and spins are usually unrelated.
At least four effects can contaminate a potential 
correlation. 
(1) As discussed in \S 2.1, 
there are large uncertainties in using the polarization angle to 
determine the spin axis; while the rotating vector model works well
in fitting polarization angles versus pulse phase in many cases,
in others it does not. Also, aberration introduces an angular shift
that depends on emission altitude, which may vary from object to object
(Blaskiewicz, Cordes \& Wasserman 1991).  ``Orthogonal'' polarization modes
for some objects are not always precisely orthogonal, raising the question
that inference of the orientation of the spin axis from the polarization
may be biased from the true value in some cases.  About half the objects
in Deshpande \etal's sample are said to be affected by orthogonal mode
flips.  
(2)  Most neutron stars may be formed in binary systems and in such cases
their present space velocities are a combination of disrupted orbital
motion and one or two kick velocities.  Any spin-axis/kick velocity
relationship can be diminished through this combination.   While it is true
that some pulsars have sufficiently large space velocities that any 
orbital contribution would be small, it is also true that 
other objects have sufficiently small space velocities that orbital motion
may have contributed significantly.
(3) Differential galactic rotation is important for distant objects
and cannot be accounted for unless the distance is known accurately
and the object has not moved far from its birth location.  Seven objects
in the Deshpande \etal sample have estimated distances $>3$ kpc and it
is possible that others originated far enough away to be so affected.  
(4) The spin-velocity angle can be altered by acceleration in the 
Galactic potential for objects older than about 10 Myr.   Five objects
in the sample exceed this age.

We stress that Deshpande \etal (1999) were careful to select objects whose
polarization angle sweeps appeared to match the shape expected from
the rotating-vector model.  However, that model has sufficient parameters
to fit any smooth position-angle curve even if that curve does not
derive from the physical conditions that underly that model.  
A cogent example is, once again, the NS-NS binary
B1913+16.  At 1.4 GHz, its pulse shape is dominated by two components
that can be associated with a `conal' beam; the position-angle curve
indeed appears to follow a curve similar to that expected from the 
rotating-vector model at the pulse phases of these components.   However,
there is also contribution to the pulse from a `core' component, as
evidenced by its appearance at lower frequencies.   In the 1.4 GHz polarization
data, the position angle curve deviates markedly from the rotating-vector
model at pulse phases corresponding to the core component
(Cordes, Wasserman \& Blaskiewicz 1990).  This raises
the suspicion that the polarization angle sweep of core components does
not map the projection of the (assumed) dipolar field lines and thus does
not allow determination of the orientation of the spin axis.  
Given that core-component emission  probably arises at or near the surface
of the neutron star (e.g. Rankin 1990), it is plausible that the magnetic
field is distorted from a dipolar form there.  Several objects in
Deshpande \etal's sample have profiles dominated by core components and 
several others show core components combined with conal emission.  

In summary, statistical studies do not allow one to conclude that the
angle between spin axes and kick directions is uniformly distributed
on the sky.  The Crab and Vela pulsars are not affected by these
uncertainties because the Chandra X-ray data give an independent
orientation vector that is presumed to coincide with their spin axes.

%%%%%%%%%%%%%%%%%%%%%%%%%%%%%%%%%%%%%%%%%%%%%%%%%%%%%%%%%%%%%%%%%%%
\section{Implications for Kick Mechanisms: Natal Kicks}

Recent studies have focused on natal kicks imparted
to the neutron star at birth. The strongest observational evidence
that kicks must be natal comes from spin-orbit misalignment 
in the binary pulsar system PSR J0045-7319 (Kaspi et al.~1996) and PSR 1913+16
(Kramer 1999) needed to produce the observed precessions. The alternative,
post-natal electromagnetic rocket mechanism (Harrison \& Tademaru 1975) 
will be discussed in \S 4.

We discuss below the implications of aligned spin-kick
in two classes of natal kick mechanisms (hydrodynamical driven
and neutrino-magnetic field driven).

\subsection{Hydrodynamically Driven Kicks}

The first class relies on hydrodynamical perturbations in core
collapse and supernova explosions. Numerical simulations indicate that local
hydrodynamical (convective) instabilities in the collapsed stellar core and its
surrounding shocked mantle (e.g., Herant et al.~1994;
Burrows et al.~1995; Janka \& M\"uller 1994,1996; Keil et al.~1996; 
Mezzacappa et al.~1998), which might in principle lead to asymmetric matter
ejection and/or asymmetric neutrino emission, are {\it not} adequate to account
for kick velocities $\go 100$~km~s$^{-1}$ (Janka \& M\"uller 1994; 
Burrows \& Hayes 1996; Keil 1998). Global asymmetric perturbations of
presupernova cores are
required to produce the observed kicks hydrodynamically (Goldreich, Lai \&
Sahrling 1996, hereafter GLS; Burrows \& Hayes 1996). GLS suggested that
overstable g-mode oscillations in the presupernova core driven by shell nuclear
burning may provide a natural seed for the initial asymmetry. Calculations
based on presupernova models of Weaver \& Woosley (1993) indicate that some
g-modes are indeed overstable 
%and may potentially grow to large amplitudes prior to core implosion 
(Lai \& Goldreich 2000b; see Lai 1999), although 
uncertainties in the presupernova models preclude a definitive prediction
of the perturbation amplitudes. Alternatively, violent convection 
in the O-Si burning shell may also produce asymmetric perturbations
(Bazan \& Arnett 1998), although it is not clear whether such
perturbations have sufficiently large scales.
During collapse, the asymmetric perturbations seeded in the outer region of the
iron core are amplified (by a factor of 5-10) by gravity (Lai \&
Goldreich 2000a).\footnote{Perturbations in the inner core (which collapses
homologously) are damped during collapse (Lai 2000), thus the proto-neutron 
star is globally spherical.} 
The enhanced asymmetric density perturbation leads to
asymmetric shock propagation and breakout, which then gives rise to asymmetry
in the explosion and a kick velocity to the neutron star (Burrows \& Hayes
1996). 

Now let us consider how rotation might affect the kick. 
%Rotation does not play a direct dynamical role in this mechanism. 
The low-order g-modes trapped in the presupernova core ($M\simeq 1.4M_\odot,
\,R\simeq 1500$~km) have periods of 1-2 seconds, much shorter than
the rotation period of the core\footnote{The minimum 
rotation period of an isolated (i.e., decoupled from the envelope) core
is about 1~s, and a real core must rotate significantly slower than this
due to core-envelope coupling. Note that even if the newly formed neutron star
rotates at the maximum breakup rate ($P_s\simeq 0.5$~ms), the corresponding
rotation period of the precollapse core is only $\simeq 11$~s.},
thus the g-modes are not affected by rotation. Also, 
since the rotational speed of the core is typically less 
than the speed of convective eddies 
($\simeq$1000-2000~km~s$^{-1}$, about $20\%$ of the sound speed) 
in the burning shell surrounding the iron core, rotation should 
not significantly affect the shell convection either\footnote{For example,
the density, temperature structure of 
the rotating presupernova stellar model of Heger, Langer \& Woosley
(2000), which does not include magnetic core-envelope coupling 
and thus overestimates the rotational
effect, is not much different from the nonrotating model.}.
(The convection zone provides an evanescent region for trapping the g-modes). 
%Therefore in general, the direction of presupernova (dipolar) asymmetry 
%is unrelated to the rotation axis. 
Under these conditions, the development of large-scale presupernova (dipolar)
asymmetry is not influenced by the core rotation.

Even though the primary thrust to the neutron star (upon core
collapse) does not depend on spin, the net kick will be
affected by rotational averaging if the asymmetry pattern
(near the shock breakout) rotates with the matter at period $P$ shorter than
the kick timescale $\tau_{\rm kick}$. 
Let $V_0$ be the kick velocity that the neutron star attains
in the case of zero rotation, and let $\theta$ be the angle between 
the primary asymmetry and the rotation axis. The expected
components of kick along the rotation axis and perpendicular to it are
(for $\tau_{\rm kick}\gg P$)
\be
V_{{\rm kick}\parallel}=V_0\cos \theta,\qquad
V_{{\rm kick}\perp}\sim {\sqrt{2}\,P\over 2\pi\,\tau_{\rm kick}}
V_0\sin \theta.
\ee
Thus the angle $\beta$ between the kick vector ${\bf V}_{\rm kick}$ and the
spin vector $\bf\Omega_s$ is given by $\tan\beta\sim 0.2(P/\tau_{\rm kick})
\tan\theta$.
%\footnote{Projected on the sky, the angle between the proper motion and
%spin axis is given by $\cos\beta_p=\cos\beta
%(\cos\theta_V\cos\theta_\Omega)^{-1}-\tan\theta_V\tan\theta_\Omega$, where 
%$\theta_V$ ($\theta_\Omega$) is the angle between ${\bf V}_{\rm kick}$
%(${\bf \Omega}$) and the plane of the sky.} 
Typically, the alignment between ${\bf V}_{\rm kick}$ and
$\bf\Omega_s$ will be achieved when $\tau_{\rm kick}\gg P$.

What is the kick timescale $\tau_{\rm kick}$? 
In the standard paradigm of core-collapse supernovae, 
the shock wave generated by core bounce
starts at a radius of $\sim 20$~km (which encloses 
$0.7M_\odot$), reaches $100-200$~km in about $20$~ms and stalls for hundreds of
milliseconds until it is revived by neutrino heating 
(perhaps enhanced by vigorous convection in the mantle) 
and then moves outward again. 
As the shock breaks out asymmetrically, momentum is imparted to the 
matter that will be incorporated into the neutron star 
on a timescale somewhat shorter than the
delay. A reasonable estimate is $\tau_{\rm kick}\sim 100$~ms 
which is the shock travel time at speed of $10^4$~km~s$^{-1}$ across $\sim 1000$~km,
the radius of the mass cut enclosing $1.4M_\odot$
(e.g., Bethe 1997; Burrows \& Hayes 1996).
%The shock breakout takes place on timescale of $\sim 100$~ms (somewhat
%shorter), thus $\tau_{\rm kick}\lo 100$~ms. 
For spin-kick alignment, the rotational period of 
the perturbation pattern at $r=100\,r_2$~km must be much less than 
$\tau_{\rm kick}$. This corresponds to a final neutron star ($R=10$~km)
spin period $P_s\ll \tau_{\rm kick}/(100\,r_2^2)\sim 1\,r_2^{-2}$~ms. 
Note that since the shock wave travels from $\sim 100$~km to 
$\sim 1000$~km while the kick is being imparted to the star
(e.g., Burrows \& Hayes 1996), $r_2=1$ corresponds to the maximum spin period
of the final neutron star.
We thus conclude that if rotation is dynamically unimportant
for the core collapse and explosion (corresponding to
$P_s\gg 1$~ms), then rotational averaging is inefficient and 
the hydrodynamical mechanism does not produce
spin-kick alignment.\footnote{In a prompt explosion,
the shock wave exits the outer core (and the mass cut boundary)
in about $20$~ms; thus $\tau_{\rm kick}\sim 20$~ms. Spin-kick 
alignment will result only when $P_s\ll 0.2 r_2^{-2}$~ms, where 
$r_2$ can be as small as $0.5$, corresponding to the radius of 
the hot proto-neutron star.}

%unless rotation plays a dynamical role, i.e., the proto-neutron star spins
%near breakup so that centrifugal forces are important, aligned kick will not
%be produced in generic situations. 

The discussion above is based on the standard picture of
core-collapse supernovae, which is valid as long as rotation does {\it not}
play a dynamically important role (other than rotational averaging)
in the supernova. If, on the other hand, rotation is dynamically important, 
the basic collapse and explosion may be qualitatively different (e.g.,
core bounce may occur at subnuclear density, the explosion is
weaker and takes the form of two-sided jets; 
M\"onchmeyer et al.~1991; Rampp, M\"uller \& Ruffert 1998; Khokhlov et
al.~1999; Fryer \& Heger 1999). 
The possibility of a kick in such 
systems has not been studied, but it is conceivable 
that an asymmetric dipolar perturbation may be coupled
to rotation, thus producing spin-kick alignment.
%Asymmetric dipolar perturbation may be coupled
%to rotation and spin-kick alignment might be possible.

\subsection{Neutrino -- Magnetic Field Driven Kicks}

The second class of kick mechanisms relies on asymmetric neutrino emission
induced by strong magnetic fields.
\footnote{Here we do not discuss kick mechanisms based on nonstandard neutrino
physics [e.g., Kusenko \& Segre (1996); Akhmedov et al.~(1997);
but see Janka \& Raffelt (1998) for critiques on these papers].}
The fractional asymmetry $\epsilon$ in the radiated
neutrino energy required to generate a kick velocity $V_{\rm kick}$ is
$\epsilon=MV_{\rm kick}c/E_{\rm tot}$ ($=0.028$ for $V_{\rm
kick}=1000$~km~s$^{-1}$, $M=1.4\,M_\odot$ and total neutrino
energy radiated $E_{\rm tot}=3\times 10^{53}$~erg). 
%As neutrinos diffuse across the proto-neutron star, they are scattered and
%absorbed/reemitted by nucleons and electrons which are slightly polarized by
%the magnetic field. 
In the magnetized nuclear medium, the neutrino scattering/absorption opacities
depend asymmetrically on the directions of neutrino momenta 
with respect to the magnetic field axis. (This is a manifestation of parity
violation in weak interactions.) Asymmetric neutrino flux can be generated 
in the outer region of the proto-neutron star (i.e., above the neutrino-matter
decoupling layer, but below the neutrinosphere) where the neutrino distribution
deviates from thermal equilibrium (Vilenkin 1995; 
Arras \& Lai 1999a,b). Arras \& Lai (1999a,b) found that averaging over all
neutrino species, the total asymmetry in neutrino flux is $\epsilon\sim
0.1\,B_{15}{\bar E_\nu}^{-2}+0.002B_{15}/T$ (the first term comes from 
the effect of quantized states of electrons in the absorption opacity, and 
the second term comes from nucleon polarization), where $B=10^{15}B_{15}$~G is
the magnetic field strength, $\bar E_\nu$ is the mean energy (in MeV)
of $\nu_e$'s and $T$ is the temperature (in MeV) in the decoupling layer.
The resulting kick velocity is $V_{\rm kick}\sim 50\,B_{15}$~km~s$^{-1}$.
Kicks of a few hundreds km~s$^{-1}$ would require the proto-neutron
star to possess an ordered component of magnetic field with magnitude
greater than $10^{15}$~G. Alternatively, since the cross section for
$\nu_e$ ($\bar\nu_e$) absorption on neutrons (protons) depends on the local
magnetic field strength, asymmetric neutrino emission can be produced if 
the field strengths at the two opposite poles of the star are different.
To generate $V_{\rm kick}\sim 300$~km~s$^{-1}$ using this mechanism
would require that the difference in the field strengths at the two opposite
stellar poles be at least $10^{16}$~G (Lai \& Qian 1998). 

A superstrong magnetic field may also play a dynamical role in the
proto-neutron star. For example, it has been suggested that the
magnetic field can induce ``dark spots'' (where the neutrino flux is
lower than average) on the stellar surface by suppressing
neutrino-driven convection (Thompson \& Duncan 1993).  While it is
difficult to quantify the kick velocity resulting from an asymmetric
distribution of dark spots, a simple estimate indicates that a local
magnetic field of at least $10^{15}$~G is needed for this effect to be
of importance.

In this second class of mechanisms, the kick is imparted to the
neutron star near the neutrinosphere\footnote{The neutrinosphere of
the proto-neutron star decreases from $\go 50$~km to near its final
value of 10-20~km in the first $100$~ms after bounce; see Burrows \&
Lattimer 1986.}  on the neutrino diffusion time, $\tau_{\rm kick}\sim
10$~seconds.  As long as the initial rotation period of the neutron
star is much less than a few seconds, spin-kick alignment is naturally
expected [see eq.~(1)].  Of course, while soft gamma repeaters and
anomalous X-ray pulsars may possess magnetic fields in excess of
$10^{14}$~G (``magnetars''; e.g., Thompson \& Duncan 1996; Vasisht \&
Gotthelf 1997; Kouveliotou et al.~1998,1999), it is unclear (and
perhaps unlikely) that radio pulsars (and Vela and Crab in particular)
had $B\go 10^{15}$~G at birth for these neutrino-magnetic field driven
kicks to be relevant.

%the internal magnetic fields of neutron
%stars and their evolution remain clouded in mystery.  

\subsection{Kick Induced Spin?}

For the two classes of kick mechanisms discussed in \S 3.1 and \S 3.2,
spin-kick alignment requires that 
the proto-neutron star have a ``primordial'' rotation 
(i.e., with angular momentum 
coming from the presupernova core and possibly at a maximum rate
if the kick is generated hydrodynamically). How about the notion that
the pulsar's spin is generated by off-centered kicks (Spruit \& Phinney 1998)?
It is certainly true that even with zero precollapse angular momentum,
some rotation can be produced in the proto-neutron star (Burrows et al.~1995
reported a rotation period of order a second generated by stochastic torques
in their 2D simulations of supernova explosions): a kick $V_{\rm kick}
=300\,V_{300}$~km~s$^{-1}$, displaced by a distance $s$ from the center,
produces a spin $\Delta\nu_s\simeq 12\,V_{300}(s/10~{\rm km})$~Hz, with the
spin axis necessarily perpendicular to the kick direction 
(Burrows et al.~1995; Spruit \& Phinney 1998). Aligned spin-kicks 
may be possible if the kick is the result of many small thrusts 
which are appropriately oriented (Spruit \& Phinney 1998) --- such 
a picture might apply if small-scale convection were responsible for the kick. 
However, numerical simulations indicate that such convection alone does not 
produce kicks of sufficient amplitude (Janka \& M\"uller 1994;
Burrows \& Hayes 1996; Keil 1998). 
As discussed in \S 3.1 and \S 3.2, kicks of a few hundreds km~s$^{-1}$ 
can be generated if global dipolar symmetry is broken (either due to
hydrodynamical perturbations or due to magnetic fields). 
%It is unlikely that aligned spin-kick can be produced generically without 
%a ``primordial'' angular momentum in the presupernova core.  

\section{Post-Natal Kick: Electromagnetic Rocket Effect}

Harrison \& Tademaru (1975) show that electromagnetic (EM) radiation 
from an off-centered rotating magnetic dipole imparts
a kick to the pulsar along its spin axis. The kick is attained
on the initial spindown timescale of the pulsar (i.e.,
this really is a gradual acceleration), and
comes at the expense of the spin kinetic energy. We have 
reexamined this effect and found that the force on the pulsar due to
asymmetric EM radiation is larger than the original 
Harrison \& Tademaru expression by a factor of four. If the dipole
is displaced by a distance $s$ from the rotation axis, and has components
$\mu_\rho,\mu_\phi,\mu_z$ (in cylindrical coordinates), the force
is given by (to leading order in $\Omega s/c$)
\be
F={8\over 15}\left({\Omega s\over c}\right){\Omega^4\mu_z\mu_\phi\over c^4}.
\ee
(The sign is such that negative $F$ implies ${\bf V}_{\rm kick}$
parallel to the spin $\bf\Omega$.) The dominant terms for the spindown 
luminosity give
\be
L={2\Omega^4\over 3c^3}\left(\mu_\rho^2+\mu_\phi^2+{2\Omega^2s^2\mu_z^2
\over 5c^2}\right).
\ee

For a ``typical'' situation, $\mu_\rho\sim\mu_\phi\sim\mu_z$, the
asymmetry parameter $\epsilon\equiv F/(L/c)$ is of order $0.4(\Omega s/c)$.
For a given $\Omega$, the maximum $\epsilon_{\rm max}=\sqrt{0.4}=0.63$ is 
achieved for $\mu_\rho/\mu_z=0$ and $\mu_\phi/\mu_z=\sqrt{0.4}\,
(\Omega s/c)$. From $M\dot V=\epsilon (L/c)=-\epsilon (I\Omega\dot\Omega)/c$,
we obtain the kick velocity
\begin{eqnarray}
&&V_{\rm kick}\simeq 260\,R_{10}^2\left({{\bar\epsilon}\over 0.1}\right)
\!\!\left({\nu_i\over 1\,{\rm kHz}}\right)^2\left[1-\left({\nu\over\nu_i}
\right)^2\right]{\rm km~s}^{-1}\nonumber\\
&&~\simeq 140 R_{10}^2\left({s\over 10\,{\rm km}}\right)
\!\!\left({\nu_i\over 1\,{\rm kHz}}\right)^3\!\left[1-\left({\nu\over\nu_i}
\right)^3\right]\!{\rm km~s}^{-1},
\label{kick1}\end{eqnarray}
where $R=10R_{10}$~km is the neutron star radius,
$\nu_i$ is the initial spin frequency, $\nu$ is the current spin frequency
of the pulsar, and
$\bar\epsilon=(\Omega_i^2-\Omega^2)^{-1}\int\!\epsilon\,d\Omega^2$;
in the second equality, we have adopted the ``typical'' situation
($\mu_\rho=\mu_\phi=\mu_z$) so that $\epsilon=0.4(\Omega s/c)$.
For the ``optimal'' condition, with $\mu_\rho=0$,
$\mu_\phi/\mu_z=\sqrt{0.4 }\,(\Omega_i s/c)$, and 
$\epsilon=\sqrt{0.4}\,
\left[2\Omega_i\Omega/(\Omega^2+\Omega_i^2)\right]$, we find
\begin{eqnarray}
&&V_{\rm kick}^{(\rm max)}\simeq
1400\,R_{10}^2\left({\nu_i\over 1\,{\rm kHz}}\right)^2\nonumber\\
&&\qquad\quad\times
\left[1-4.66\left({\nu\over\nu_i}-\tan^{-1}{\nu\over\nu_i}\right)\right]
{\rm km~s}^{-1}.
\label{kickmax}\end{eqnarray}
Thus if the neutron star was born rotating at $\nu_i\go 1$~kHz, 
it is possible, in principle, to generate spin-aligned kick
of a few hundreds km~s$^{-1}$ or even 1000~km~s$^{-1}$.

Equations (\ref{kick1}) and (\ref{kickmax}) assume that the rotational energy
of the pulsar entirely goes to electromagnetic radiation. Recent
work has shown that a rapidly rotating ($\nu\go 100$~Hz)
neutron star can potentially lose significant angular momentum to 
gravitational waves generated by unstable r-mode oscillations
(e.g., Andersson 1998; Lindblom, Owen \& Morsink 1998;
Owen et al.~1998; Andersson, Kokkotas \& Schutz 1999; Ho \& Lai 2000). 
If gravitational radiation carries away the rotational energy of the neutron
star faster than the EM radiation does, then the electromagnetic rocket effect
will be much diminished\footnote{Gravitational radiation can also carry away
linear momentum, but the effect for a neutron star is negligible.}.
In the linear regime, the r-mode amplitude $\alpha\sim \xi/R$ (where 
$\xi$ is the surface Lagrangian displacement; see the references cited above
for more precise definition of $\alpha$) grows due to gravitational radiation
reaction on a timescale $t_{\rm grow}\simeq 19\,(\nu/1~{\rm kHz})^{-6}$~s.
Starting from an initial amplitude $\alpha_i\ll 1$, the mode grows
to a saturation level $\alpha_{\rm sat}$ in time $t_{\rm grow}\ln(\alpha_{\rm
sat}/\alpha_i)$ during which very little rotational energy is lost. 
After saturation, the neutron star spins down due to gravitational radiation
on a timescale
\be
\tau_{\rm GR}=\left|{\nu\over\dot\nu}\right|_{\rm GR}\simeq
100\,\alpha_{\rm sat}^{-2}\left({\nu\over 1\,{\rm kHz}}\right)^{-6}\,{\rm s},
\ee
(Owen et al.~1998). By contrast, the spindown time due to EM radiation alone is
\be
\tau_{\rm EM}=\left|{\nu\over\dot\nu}\right|_{\rm EM}
\simeq 10^7\,B_{13}^{-2}\left({\nu\over 1\,{\rm
kHz}}\right)^{-2}\,{\rm s},
\label{taukick}\ee
where $B_{13}$ is the surface dipole magnetic field in units of $10^{13}$~G.
Including gravitational radiation, the kick velocity becomes
\begin{eqnarray}
&&V_{\rm kick}={I\over Mc}\int_{\Omega}^{\Omega_i}\!\!\epsilon\,\,
{\tau_{\rm GR}\over\tau_{\rm GR}+\tau_{\rm EM}}\,\Omega d\Omega\nonumber\\
&&\quad\quad \,\simeq 260\,R_{10}^2\left({{\bar\epsilon}\over 0.1}\right)
\!\left({\nu_i\over 1\,{\rm kHz}}\right)^2\nonumber\\
&&\qquad\quad\times{1\over \beta}\ln\left[{1+\beta\over
1+\beta\,(\nu/\nu_i)^2}\right]
{\rm km~s}^{-1},
\label{kick2}\end{eqnarray}
where in the second equality we have replaced $\epsilon$ by
constant mean value $\bar\epsilon$, and $\beta$ is defined by 
\be
\beta\equiv \left({\tau_{\rm EM}\over\tau_{\rm GR}}\right)_i\simeq
\left({\alpha_{\rm sat}\over 10^{-2.5}}\right)^{2}\left({\nu_i\over 1\,{\rm
kHz}}\right)^{4}B_{13}^{-2}.
\ee
For $\beta\ll 1$, equation (\ref{kick2}) becomes (\ref{kick1}); for
$\beta\gg 1$, the kick is reduced by a factor $1/\beta$. 

Clearly, for the EM rocket to be viable as a kick mechanism
at all requires $\beta\lo 1$. The value of $\alpha_{\rm sat}$ is
unknown. Analogy with secularly unstable bar-mode in a Maclaurin spheroid 
implies that $\alpha_{\rm sat}\sim 1$ is possible. It has been
suggested that turbulent dissipation in the boundary layer near the crust
(if it exists early in the neutron star's history) may limit 
$\alpha_{\rm sat}$ to a small value of order $10^{-2}$-$10^{-3}$ (Wu, Matzner
\& Arras 2000). The theoretical situation is not clear at this point.

The EM rocket effect has not been popular in recent years because
empirical tests using radio polarization and proper motion have not
yielded a positive correlation and because
it is widely thought that radio pulsars are born with spin periods
of 0.02-0.5~s (e.g., Lorimer et al.~1993). However,
as discussed in \S 2.3, the polarization-proper motion
correlation might not be detectable even if kicks are always aligned
with spin axes. The recent discovery 
of the 16~ms X-ray pulsar (PSR J0537-6910) associated with the
Crab-like supernova remnant N157B (Marshall et al.~1998)
%in the Large Magellanic Cloud 
implies that at least some neutron stars are born with spin periods in
the millisecond range. In particular, a recent analysis of the
energetics of the Crab Nebula suggests an initial spin period $\sim
3-5$ ms followed by fast spindown on a time scale of 30 yr (Atoyan 1999).
As for the Vela pulsar, the energetics of the remnant do not yield an
unambiguous constraint on the initial spin (S. Reynolds, private
communication).  Thus it seems prudent to consider the EM rocket
effect as a possible kick mechanism. We do note, however, that such
slow post-natal kick may have difficulty explaining some properties of
binary pulsar systems, such as the spin-orbit misalignment in PSR
J0045-7319 (Kaspi et al.~1996) and PSR 1913+16 (Kramer 1999) needed to
produce the observed precessions.
\footnote{For instance, in the case of PSR J0045-7319 -- B star binary:
if we assume that the orbital angular momentum of the 
presupernova binary is aligned with the spin of the B star, then the current
spin-orbit misalignment can only be explained by a fast
kick with $\tau_{\rm kick}$ less than the post-explosion orbital period 
$P_{\rm orb}$; if we further assume that the iron core of the progenitor of
PSR J0045-7319
had a spin aligned with the orbit, then the fast kick must not have
been along the neutron star spin axis. See also \S 2.3 for the
case of PSR B1913+16.} 
Similarly, a slow kick (with $\tau_{\rm kick}\go P_{\rm orb}$)
may be inconsistent with the neutron star binary populations (e.g.,
Dewey \& Cordes 1987; Fryer \& Kalogera 1997; Fryer et al.~1998).  
Some of these issues will be addressed in a future paper.

\section{Conclusion}

Motivated by the apparent alignment between the spin axis and the
proper motion for the Vela and Crab pulsars 
as revealed by recent Chandra observations 
and by analysis of radio polarization (\S 2), we have examined in this paper
the questions of how rotation affects kick and whether spin-kick alignment is
generally expected in different classes of theoretical models of neutron star
kicks. We find that 

(1) For hydrodynamically driven kicks, spin-kick alignment is possible
only if rotation plays a dynamically important role in the core
collapse and explosion, corresponding to neutron star initial spin period 
$P\lo 1$~ms (\S 3.1); 

(2) For neutrino-magnetic field driven kicks, spin-kick alignment is
almost always produced (\S 3.2); but for these class of mechanisms to be 
relevant, the proto-neutron star must have magnetic field stronger than
$10^{15}$~G; 

(3) Post-natal kicks due to the electromagnetic rocket effect always lead to 
spin-kick alignment; even with our upward revised (by a factor of 4)
force expression compared to the original Harrison-Tademaru formula,  
for this mechanism to be of interest, the
proto-neutron star must rotate at frequency $\go 1$~kHz, and
gravitational radiation (via r-mode instability) driven spindown
must be less efficient than magnetic braking --- this puts a constraint
on the magnetic field and r-mode amplitude in the young neutron star
(\S 4). 

Regardless of which kick mechanism is at work, if spin-kick
alignment turns out to be a generic feature, then neutron stars must have
non-negligible (and perhaps maximum) initial rotation. Theoretically,
it has been argued that efficient core-envelope coupling via magnetic
stress in the presupernova star results in negligible ``primordial''
angular momentum for the neutron star (Spruit \& Phinney 1998), although 
there are different views (e.g., Livio \& Pringle 1998). If, on the 
other hand, future observations show that spin-kick alignment is 
not a generic feature, then we must conclude that the alignment in Vela and
Crab pulsars is a coincidence, and the origin of neutron star kicks 
lies in hydrodynamics.

\acknowledgments
This work is supported in part
by NASA grants NAG 5-8356 and NAG 5-8484, by NSF grants AST 9819931 and 
AST 9986740, and by a research fellowship (to D.L.) from the Alfred P. Sloan 
foundation.

%%%%%%%%%%%%%%%%%%%%%%%%%%%%%%%%%%%%%%%%%%%%%%%%%%%%%%%%%%%%%%%%%%%%%

\end{document}